\documentclass[aps,prl,twocolumn,showpacs,amsmath,amssymb,floatfix,superscriptaddress]{revtex4}
\usepackage{graphicx}
\usepackage{Jonasmacros}
\usepackage{bm}
\usepackage{fancyheadings}
\newcommand{\beqa}{\begin{eqnarray}}
\newcommand{\eeqa}{\end{eqnarray}}

\begin{document}
\preprint{}
\title{Spin Dynamics in a Tunnel Junction between Ferromagnets}
\author{Jonas Fransson}
\email{jkfransson@gmail.com}
\affiliation{Theoretical Division, Los Alamos National Laboratory,
Los Alamos, New Mexico 87545 }
\affiliation{Center for Nonlinear Studies, Los Alamos National Laboratory,
Los Alamos, New Mexico 87545 }
\author{Jian-Xin Zhu}
\email{jxzhu@lanl.gov}
\homepage{http://theory.lanl.gov}
\affiliation{Theoretical Division, Los Alamos National Laboratory,
Los Alamos, New Mexico 87545 }

\begin{abstract}
The dynamics of a single spin embedded in the tunnel junction (quantum point contact) between ferromagnets is addressed. Using the Keldysh technique, we derive a quantum Langevin equation. As a consequence of the spin-polarization in the leads, the spin displays a rich and unusual dynamics. Parallel configured and equally strong magnetic moments in the leads yield an ordinary spin precession with a Larmor frequency given by the effective magnetic field.
Unequal and/or non-parallel configured magnetization, however, causes nutation of the spin in addition to the precession. Our predictions may be directly tested for macroscopic spin clusters.
\end{abstract}
\date{\today}
\maketitle

The interest in a number of techniques that allow one to detect and manipulate a single spin in the solid state remains tremendous both experimentally~\cite{Mana89,Mana00,Durkan02,Manoharan02} and 
theoretically~\cite{Engel01,Balatsky02a,Balatsky02b,Mozyrsky02,Zhu02,Bulaevskii03,Hastings04}. Being a crucial element in spintronics and spin-based quantum information processing, such studies are also of fundamental importance. So far, most of efforts~\cite{Balatsky02a,Mozyrsky02,Zhu02,Bulaevskii03,Hastings04} have been focused on understanding the mechanism for the tunneling current modulation, which is the hallmark of a single spin detection by using  the scanning tunneling microscopy (STM)~\cite{Mana89,Mana00,Durkan02}. 
Recently,  the coupling between a single spin and supercurrent in Josephson junctions has also been studied~\cite{Zhu04,Bulaevskii04}. 

Especially, a single spin nutation induced by an {\em ac} supercurrent in a {\em dc} biased Josephson junction was shown for the first time in Ref.~\cite{Zhu04}. From the view point of single spin manipulation, such kind of nutation provides a significant implication for the control of spin dynamics {\em electrically}. The question is whether the spin dynamics (or even switching) can be realized  in other types (i.e., non-superconducting) of leads, which are easily accessible experimentally. Recently we have examined this issue in normal conducting leads~\cite{Zhu06}.  It was found that the spin-flip process of tunneling electrons is important to manipulate the spin and the flip-rate determines the efficiency of spin manipulation. Studies of local spin-dynamics in quantum dots between ferromagnetic leads were recently reported \cite{gurvitz2005,braun2006}, however, these studies were concerned with the time-dependent effects and the noise of a local spin under the influence of stationary external fields.

In this paper, we study the spin dynamics of a single spin embedded in a tunneling junction between two ferromagnetic leads. A quantum Langevin equation is derived for the single spin dynamics. Through the resulting equation we show that the tunneling between the ferromagnetic leads converts the electric field, e.g. bias voltage, into an effective magnetic field. The resulting effective magnetic field, however, depends on the relative orientation between the magnetization in the two leads. Parallel (i.e., FM-type)
alignment of equally strong magnetization in the two leads yields a shift of the Larmor frequency, which scales as the squared ratio between the induced and the external magnetic fields. The FM-type alignment but with unequal magnetization or anti-parallel (AFM-type) alignment in the two leads, on the other hand, leads to nutations in the precession of the spin about the external field.

The model system we consider consists of two ferromagnetic leads coupled to each other by a single spin $\bfS$. We assume that the tunnel junction is formed by a quantum point contact between the leads, so that the magnetic fields generated at the tips of the magnetic leads can be neglected. We also neglect the direct interaction of the spin with the two leads. Then, the system Hamiltonian can be 
written as
\begin{equation}
\Hamil=\Hamil_L+\Hamil_R+\Hamil_S+\Hamil_T.
\label{eq-system}
\end{equation}
The first two terms $\Hamil_{L/R}=\sum_{k\sigma\in L/R}\leade{k}\cdagger{k}\cc{k}$ describe the electrons in the leads, where an electron is created (annihilated) in the left/right ($L/R$) lead at the energy $\leade{k}$ by $\cdagger{k}\ (\cc{k})$. Henceforth, we assign subscripts $p\ (q)$ to electrons in the left (right) lead. The Hamiltonian for a free spin $\bfS$ in the presence of a magnetic field $\bfB$ is given by
\begin{equation}
\Hamil_S=-g\mu_B\bfB\cdot\bfS,
\label{eq-HS}
\end{equation}
where $g$ and $\mu_B$ are the gyromagnetic ratio and Bohr magneton, respectively. The two leads are weakly coupled via the tunneling Hamiltonian
\begin{equation}
\Hamil_T=\sum_{pq\alpha\beta}\left(\csdagger{p\alpha}
	[T_{\alpha}\delta_{\alpha\beta}
		+T_1\bfS\cdot\bm{\sigma}_{\alpha\beta}]\cs{q\beta}+H.c.\right).
\label{eq-HT}
\end{equation}
Here, $\bm{\sigma}_{\alpha\beta}$ is the vector of Pauli spin matrices, with spin indices $\alpha,\beta$, whereas $T_\alpha$ is the direct tunneling rate between the spin-polarized  leads, and $T_1$ is the tunneling rate between the leads modulated by the local spin. In this model we neglect spin-flip transitions in the direct tunneling between the leads. 
For convenience we take the respective amplitudes to be momentum independent (although it is not required). Typically, from the expansion of the work function for tunneling $T_1/T_\alpha \sim J/U$ \cite{Balatsky02b}, where $J$ is the exchange coupling strength between the transport electrons and the local spin and $U$ is a spin-independent tunneling barrier. We further allow a weak external magnetic field $B \sim 10^{2}-10^{4}\; \mbox{Gauss}$, which is applied along the $z$-direction in the $z$-$x$ plane perpendicular to the electron tunneling direction ($y$ axis).

When a time-dependent voltage bias is applied across the tunneling barrier, such that $V(t)=V_{dc}+V_{ac}\cos(\omega_{0}t)$, where $V_{dc}$ and $V_{ac}$ are the dc and ac components, and $\omega_{0}$ is the frequency of the ac field, a dipole will be formed around the barrier region through the accumulation or depletion of electron charge. This process results in the time dependence of single-particle energies: $E_p=\epsilon_p+W_L(t)$ and $E_q=\epsilon_q+W_R(t)$, with the constraint $W_L(t)-W_R(t)=eV_{ac} \cos(\omega_0t)$. However, the occupation of each state in the respective contact remains unchanged and is determined by the distribution established before the time dependence is turned on. Therefore, the chemical potentials on the left $\mu_{L}$ and on the right lead $\mu_{R}$ differs by the dc component of the applied voltage bias, $\mu_{L}-\mu_{R}=eV_{dc}$. The tunneling junction with the spin then has two time scales: The Larmor precession frequency of the spin $\omega_L = g \mu_B B$ and the characteristic frequency $\omega_{0}$ of the ac field.

The model is gauge transformed by
\begin{equation}
\hat{U}=e^{-i\int_{t_0}^t[\mu_L+W_L(t')]N_Ldt'}e^{-i\int_{t_0}^t[\mu_R+W_R(t')]N_Rdt'},
\label{eq-gauge}
\end{equation}
with $N_{L/R}=\sum_{k\sigma\in L/R}\cdagger{k}\cc{k}$, and we thus obtain the model $K=K_L+K_R+K_S+K_T$, where $K_{L/R}=\sum_{k\sigma\in L/R}\xi_{k\sigma}\cdagger{k}\cc{k}$, $\xi_{k\sigma}=\leade{k}-\mu_{L/R}$, $K_S=\Hamil_S$, and
\begin{equation}
K_T=\sum_{pq\alpha\beta}\left(\csdagger{p\alpha}
	\hat{T}_{\alpha\beta}\cs{q\beta}e^{i\phi(t)}+H.c.\right).
\label{eq-KT}
\end{equation}
Here, we have introduced the notation $\hat{T}_{\alpha\beta}(t)=T_\alpha\delta_{\alpha\beta}+T_1\bfS(t)\cdot\bm{\sigma}_{\alpha\beta}$ and 
$\phi(t)=e\int_{t_0}^t[V_{dc}+V_{ac}\cos{\omega_0t'}]dt'$.

We now derive the effective action via the Keldysh technique \cite{keldysh1965}. If all external fields are the same on both forward and backward branches of the Keldysh contour ($C$), then the partition function ${\cal Z}=\tr{{\rm T}_C \exp{[-i\oint_C K_T(t)dt]}}=1$, where the trace runs over both electron and spin degrees of freedom. We take the partial trace in ${\cal Z}$ over the lead electrons (bath) to obtain an effective spin action. In the present situation, this action represents the interaction of the magnetic spin with a non-equilibrium environment. The tunneling contribution to the resulting spin action reads $i\delta{\cal S}=-(1/2)\oint_C\oint_C\av{{\rm T}_CK_T(\bfS(t),t)K_T(\bfS(t'),t')}dtdt'$, much in the spirit of Refs. \cite{AES_LO}.

For brevity we put $A_{\alpha\beta}=\sum_{pq}\csdagger{p\alpha}\cs{q\beta}$. The tunneling Hamiltonian of a voltage biased junction then reads
\begin{equation}
K_T(\bfS(t),t)=\sum_{\alpha\beta}(\hat{T}_{\alpha\beta}(t)A_{\alpha\beta}e^{i\phi(t)}+H.c.).
\label{eq-KTS}
\end{equation}
For magnetic leads, the correction to the effective action for the spin dynamics is thus given by
\begin{eqnarray}
i\delta{\cal S}=-i\oint_C\oint_C
	\hat{T}_{\alpha\beta}(t){\cal D}_{\alpha\beta}(t,t')\hat{T}_{\beta\alpha}(t')
	e^{i[\phi(t)-\phi(t')]}
	dtdt'\;,
\label{eq-ids}
\end{eqnarray}
where ${\cal D}_{\alpha\beta}(t,t')=-i\av{{\rm T}_CA_{\alpha\beta}(t)A_{\alpha\beta}^\dagger(t')}$.

Performing standard Keldysh manipulations, defining upper and lower spin fields $\bfS^{u/l}$ residing on the forward/backward contours and reducing the time ordered integral over Keldysh contour to the integral over forward running time at the cost of making the Green functions (GF) ${\cal D}_{\alpha\beta}$ a $2\times2$ matrix, for each spin combination $\alpha\beta$. We then perform a rotation to the classical and quantum components
\begin{equation}
\bfS^c\equiv(\bfS^u+\bfS^l)/2,\ \bfS^q\equiv\bfS^u-\bfS^l,\ \bfS^c\cdot\bfS^q=0,
\label{eq-S}
\end{equation}
which makes the matrix Green function uniquely determined in terms of the retarded/advanced component ${\cal D}_{\alpha\beta}^{r/a}(t,t')=\mp i\theta(\pm t\mp t')\av{\com{A_{\alpha\beta}(t)}{A_{\alpha\beta}^\dagger(t')}}$, and the Keldysh component ${\cal D}_{\alpha\beta}^K(t,t')=-i\av{\anticom{A_{\alpha\beta}(t)}{A_{\alpha\beta}^\dagger(t')}}$. The procedure leads to $\delta{\cal S}=\delta{\cal S}_c+\delta{\cal S}_q$, where
\begin{equation}
\delta{\cal S}_c=\int[S_z^q(t)K_z^{(1)}(t,t')+S_i^q(t)K_{ij}^{(2)}(t,t')S_j^c(t')]
	dtdt'
\label{eq-dsc}
\end{equation}
and
\begin{equation}
\delta{\cal S}_q=
	\int S_i^q(t)K_{ij}^{(3)}(t,t')S_j^q(t')dtdt',
\label{eq-ds}
\end{equation}
where summation over repeated indices $i,j=x,y,z$, is understood. Here the kernels are given by
\begin{eqnarray}
K_z^{(1)}(t,t')&=&
	-T_1\sum_\sigma\sigma_{\sigma\sigma}^zT_\sigma
		[{\cal D}^r_{\sigma\sigma}(t,t')e^{i[\phi(t)-\phi(t')]}
\nonumber\\&&
		+{\cal D}^a_{\sigma\sigma}(t',t)e^{-i[\phi(t)-\phi(t')]}]
\label{eq-K1}\\&&\hspace{-1cm}=
	-2T_1\theta(t-t')\sum_{pq\sigma}\sigma_{\sigma\sigma}^zT_\sigma
		[f(\xi_{p\sigma})-f(\xi_{q\sigma})]
\nonumber\\&&\hspace{-0.5cm}\times
		\sin{[(\xi_{p\sigma}-\xi_{q\sigma})(t-t')+\phi(t)-\phi(t')]}  \;,
\nonumber\\
K_{ij}^{(2)}(t,t')&=&
	-T_1^2\sum_{\alpha\beta}\sigma_{\alpha\beta}^i
		[{\cal D}^r_{\alpha\beta}(t,t')e^{i[\phi(t)-\phi(t')]}
\nonumber\\&&
		+{\cal D}^a_{\beta\alpha}(t',t)e^{-i[\phi(t)-\phi(t')]}]
		\sigma_{\beta\alpha}^j
\label{eq-K2}\\&&\hspace{-1cm}=
	iT_1^2\theta(t-t')\sum_{pq\alpha\beta}\sigma_{\alpha\beta}^i
	\{[f(\xi_{p\alpha})-f(\xi_{q\beta})]
\nonumber\\&&\hspace{-0.5cm}\times
	e^{i(\xi_{p\alpha}-\xi_{q\beta})(t-t')
		+i[\phi(t)-\phi(t')]}
\nonumber\\&&\hspace{-0.5cm}
	-[f(\xi_{p\beta})-f(\xi_{q\alpha})]
\nonumber\\&&\hspace{-0.5cm}\times
	e^{-i(\xi_{p\beta}-\xi_{q\alpha})(t-t')
		-i[\phi(t)-\phi(t')]}\}\sigma_{\beta\alpha}^j\;,
\nonumber\\
K_{ij}^{(3)}(t,t')&=&
	-\frac{T_1^2}{2}\sum_{\alpha\beta}\sigma_{\alpha\beta}^i
		{\cal D}_{\alpha\beta}^K(t,t')\sigma_{\beta\alpha}^j
		e^{i[\phi(t)-\phi(t')]}
\label{eq-K3}\\&&\hspace{-1cm}=
	i\frac{T_1^2}{2}\sum_{pq\alpha\beta}\sigma_{\alpha\beta}^i\sigma_{\beta\alpha}^j
	[f(\xi_{p\alpha})+f(\xi_{q\beta})
\nonumber\\&&\hspace{-0.5cm}
	-2f(\xi_{p\alpha})f(\xi_{q\beta})]
	e^{i(\xi_{p\alpha}-\xi_{q\beta})(t-t')+i[\phi(t)-\phi(t')]}\;.
\nonumber
\end{eqnarray}
Here $f(\xi)$ is the Fermi distribution function $f(\xi) = 1/[\exp(\xi/k_{B}T)+1]$. 
It follows that the only non-vanishing components of $K_{ij}^{(2)}$ are $K_{yy}^{(2)}=K_{xx}^{(2)}$, $K_{zz}^{(2)}$, and $K_{yx}^{(2)}=-K_{xy}^{(2)}$.

To properly describe the dynamics of the spin, we employ the path integral representation for the spin fields. So in addition to the terms, $-\oint \Hamil_{S}(t) dt$, the action for a free spin also contains a Wess-Zumino-Witten-Novikov (WZWN), $\mathcal{S}_{WZWN}$, which describes the Berry phase accumulated by the spin as a result of motion of the spin on the sphere. We generalize this action for nonequlibrium dynamics within the Keldysh contour formalism, which can be expressed as~\cite{Zhu04}
\begin{eqnarray}
\mathcal{S}_{WZWN} = \frac{1}{S} \int dt\, \mathbf{S}^{q} \cdot
(\mathbf{S}^c \times \partial_{t} \mathbf{S}^c).
\end{eqnarray}
The total effective spin action is given by:
\begin{eqnarray}
\mathcal{S}_\text{eff} = \mathcal{S}_{WZWN}  + g \mu_{B} \int dt
\mathbf{B} \cdot \mathbf{S}^{q}(t) + \delta \mathcal{S}_c
+\delta\mathcal{S}_{q}\;. \label{EQ:WZWN0}
\end{eqnarray}
As seen from Eqs.~(\ref{eq-K1})-(\ref{eq-K3}), the first three
terms on the right hand side of Eq.~(\ref{EQ:WZWN0}) are
real, which determine the quasi-classical equation of motion,
while $\delta\mathcal{S}_{q}$ is imaginary, which stands for the
fluctuations of the spin field $\mathbf{S}^{q}$. This means that
the quantum effects have indeed been included even in the
semi-classical approximation. We perform the Hubbard-Stratonovich
transformation with an auxiliary stochastic field $\bm{\xi}(t)$ to
decouple the quadratic term in $\delta{\cal S}_{q}$.  The total
effective action is rewritten as:
\begin{eqnarray}
\lefteqn{
\mathcal{S}_\text{eff}=\mathcal{S}_{WZWN}+g \mu_{B} \int
[\mathbf{B}+\bm{\xi}(t)] \cdot \mathbf{S}^{q}(t)dt
}
\label{eq-Seff}\\&&
	+\int[S_z^q(t)K^{(1)}_z(t,t')+S^{q}_i(t)K_{ij}^{(2)}(t,t')S^c_j(t')]dtdt',
\nonumber
\end{eqnarray}
where the fluctuating random magnetic fields satisfy the correlation functions
\begin{equation}
(g\mu_{B})^{2}\langle\xi_i(t)\xi_j(t^{\prime})\rangle
	=-2iK_{ij}^{(3)}(t,t^{\prime}).
\label{eq-fluctuating}
\end{equation}

Equation~(\ref{eq-Seff}) constitutes the central formula for the following analysis. Notice that 
the kernel $K_{ij}^{(2)}$, which connects the quantum and classical spin fields (see Eq.~(\ref{eq-Seff})), 
represents the effects of electron degrees of freedom.  It  has the following approximate behavior in time: 
\begin{equation} 
K_{ij}^{(2)}(t,t^{\prime}) \propto \frac{\cos[\mu(t-t^{\prime})]}{[(t-t^{\prime}) + i\eta]^{2}}\;,
\label{EQ:asymptotic}
\end{equation}
where $\eta$ is an infinitesimal and $\mu$ is the energy scale at the order of voltage bias. 
A similar type of approximate behavior has also been observed in the case of a mechanical oscillator 
coupled to two Luttinger liquids~\cite{MBHastings03}. Equation~(\ref{EQ:asymptotic}) suggests that the kernel $K_{ij}^{(2)}$ is peaked at $t-t^{\prime}=0$ while oscillates at larger times. The characteristic time is set by $1/\mu$, which also measures the width of the peak around $t-t^{\prime}=0$. 
Because of the oscillating nature of  $K_{ij}^{(2)}$ at the large time scale, 
the corresponding time integral on the right-hand side of Eq.~(\ref{eq-Seff}) is dominated from the 
time range of $1/\mu$ around $t-t^{\prime}=0$. 
When $1/\mu \ll 1/\omega_{L}$ (i.e., $\mu \gg \omega_{L}$), we are in the regime where 
the spin dynamic processes are much slower  than those of the conduction electrons. Since the energy associated with the spin dynamics, $\hbar\omega_L\sim1 \mu$eV,  
while  the typical energy for the electronic degrees of freedom at the  order of 1 meV is routine, the above regime is easily accessible.  
Under this condition, it is reasonable for us  to use the approximation $\bfS^c(t')\approx\bfS^c(t)+(t'-t)d\bfS^c(t)/dt$. The variational equations $\delta{\cal S}_\text{eff}/\delta\bfS^q(t)=0$ then yield
\begin{equation}
\frac{d{\bf n}}{dt}=\alpha(t)\frac{d{\bf n}}{dt}\times{\bf n}
	+g\mu_B{\bf n}\times[\bfB_\text{eff}(t)+\bm{\xi}(t)] \;,
\label{eq-Seq}
\end{equation}
where we have put $\bfS^c(t)=S{\bf n}(t)$, whereas $\alpha(t)=S\int\mathbb{K}^{(2)}(t,t')(t-t')dt'$, with the $3\times3$ matrix  $\mathbb{K}^{(2)}(t,t')=\{K_{ij}^{(2)}(t,t')\}_{i,j=x,y,z}$, and $\bfB_\text{eff}(t)=\bfB(t)+\bfB_\text{ind}^{(1)}(t)+\bfB_\text{ind}^{(2)}(t)$. We expect that the Langevin term ($\bm{\xi}(t)$) is suppressed at frequencies much lower than the exchange interaction field in the leads.

First we note that $\alpha(t)\sim ST_1^2\omega_0/D$ at zero temperature, where $2D$ is the band-width in the leads. Hence, for large $D$, which is reasonable for metals, $\alpha$ is negligibly small such that first term to the right in Eq. (\ref{eq-Seq}) drops out. It is reasonable to believe that this would be true also for finite temperatures. This result is different from the case of a dc biased superconducting tunnel junction, where $\alpha(t)$ is finite leading to spin nutation~\cite{Zhu04}.

The magnetic leads induce the field $\bfB_\text{ind}^{(1)}=B_\text{ind}^{(1)}\hat{{\bf z}}$  with
\begin{eqnarray}
\lefteqn{
g\mu_BB_\text{ind}^{(1)}(t)=\int K_z^{(1)}(t,t')dt'
}
\nonumber\\&=&
	-2\pi T_1\sum_\sigma\sigma_{\sigma\sigma}^zT_\sigma N_{L\sigma}N_{R\sigma}
		\sum_{nm}J_n(eV_{ac}/\omega_0)
\nonumber\\&&\times
		J_m(eV_{ac}/\omega_0)
		(eV_{dc}+m\omega_0)\sin{\omega_0(n-m)t},
\label{eq-Kz}
\end{eqnarray}
where $N_{L\sigma/R\sigma}$ is the density of states of the spin $\sigma$ sub-band in the left/right lead, whereas $J_n(x)$ is the $n$th Bessel function. This induced magnetic field contributes to the spin dynamics whenever, at least, one of the leads is spin-polarized.

The second induced magnetic field, $\bfB_\text{ind}^{(2)}(t)$ is defined such that ${\bf n}\times g\mu_B\bfB_\text{ind}^{(2)}(t)={\bf n}\times S\int\mathbb{K}^{(2)}(t,t')dt'{\bf n}$, and we find that this field can be written as 
\begin{eqnarray}
g\mu_B\bfB_\text{ind}^{(2)}(t)&=&
	S\int\Bigl[K_{xy}^{(2)}(t,t')(n_y\hat{{\bf x}}-n_x\hat{{\bf y}})
\nonumber\\&&
		-[K_{xx}^{(2)}(t,t')-K_{zz}^{(2)}(t,t')]n_z\hat{{\bf z}}\Bigr]dt'.
\label{eq-B'}
\end{eqnarray}
Here, the longitudinal component is proportional to
\begin{eqnarray}
\lefteqn{
\int[K_{xx}^{(2)}(t,t')-K_{zz}^{(2)}(t,t')]dt'=
}
\nonumber\\&=&
	2\pi T_1^2(N_{L\up}-N_{L\down})(N_{R\up}-N_{R\down})
		\sum_{nm}J_n(eV_{ac}/\omega_0)
\nonumber\\&&\times
		J_m(eV_{ac}/\omega_0)
		(eV_{dc}+m\omega_0)\sin{\omega_0(n-m)t},
\label{eq-Kx-z}
\end{eqnarray}
which thus contributes to the spin dynamics whenever both leads are spin-polarized. The transverse component of the induced field, which is proportional to 
\begin{eqnarray}
\lefteqn{
\int K_{xy}^{(2)}(t,t')dt'=
}
\nonumber\\&=&
	-2\pi T_1^2(N_{L\up}N_{R\down}-N_{L\down}N_{R\up})
		\sum_{nm}J_n(eV_{ac}/\omega_0)
\nonumber\\&&\times
		J_m(eV_{ac}/\omega_0)
		(eV_{dc}+m\omega_0)\cos{\omega_0(n-m)t},
\label{eq-Kxy}
\end{eqnarray}
vanishes for magnetic leads whenever $N_{L\up}N_{R\down}=N_{L\down}N_{R\up}$, that is, when the magnetizations of the leads are equal and in parallel configuration. Thus, under those conditions, the resulting motion of the spin will be influenced by the magnetizations of the leads. More interestingly though, whenever the magnetizations of the leads are non-equal and/or in a non-parallel configuration (including anti-parallel) we expect the spin dynamics to become significantly modified, since then the equation of motion~(\ref{eq-Seq}), contains also a non-trivial $z$-component which is induced by the non-vanishing transverse component of the induced magnetic field.

The contributions expressed in Eqs. (\ref{eq-Kz})-(\ref{eq-Kxy}) are caused by the spin imbalance, or spin-polarization, in the leads and describe the effect of the current flow on the local spin dynamics. The field $\bfB_\text{ind}^{(1)}$ arise whenever there is a spin imbalance in, at least, one of the lead. By its presence the local spin exerts a precession about its local direction. The $z$-contribution of the field $\bfB_\text{ind}^{(2)}$, given in Eq. (\ref{eq-Kx-z}), is finite only when both leads are magnetic, and its effect on the local spin is analogous to that of $\bfB_\text{ind}^{(1)}$. The transverse contribution from $\bfB_\text{ind}^{(1)}$, given in Eq. (\ref{eq-Kxy}), which is of main interest in this paper, requires that the magnetic moments of the leads are non-parallel and gives maximal effect when they are anti-ferromagnetically (AFM) aligned. Caused by the misaligned magnetic moments of the leads, the tunneling electron has to undergo spin-flip transitions in order to tunnel between the leads. The spin-flipping tunneling electrons tend to change the local magnetic field in the vicinity of the local spin, such that its local direction is slightly altered. In the case with AFM aligned leads, while the local spin tends to line up with the magnetic moment of the source lead, an ac component in the bias voltage tends to cause wobbling of the spin, as is shown below.

The classical equation of motion should be consistent with the parametrization of the spin on the unit sphere, $S{\bf n}=S(\cos{\phi}\sin{\theta},\sin{\phi}\sin{\theta},\cos{\theta})$. Letting the external magnetic field $\bfB$ be oriented along the $z$-axis, we find the equations 
\begin{equation}
\begin{array}{rcl}
\displaystyle \frac{d\phi}{dt} & = &
	\displaystyle - g\mu_B[B(t)+B_\text{ind}^{(1)}(t)]
\\ &&
	\displaystyle +S\int[K_{xx}^{(2)}(t,t')-K_{zz}^{(2)}(t,t')]dt'\cos{\theta},
\\\\
\displaystyle \frac{d\theta}{dt} & = &
	\displaystyle S\int K_{xy}^{(2)}(t,t')dt'\sin{\theta}.
\end{array}
\label{eq-angles}
\end{equation}
Defining $A_{nm}(V_{dc},V_{ac})$ such that $S\int K_{xy}^{(2)}(t,t')dt'=\sum_{nm}A_{nm}(V_{dc},V_{ac})\cos{\omega_0(n-m)t}$, and letting the spin initially at time $t_0=0$ be oriented at an angle $\theta_0$ relative to $\bfB$ we explicitly find the polar coordinate
\begin{eqnarray*}
\theta(t)&=&2\arctan{\biggl[\tan\frac{\theta_0}{2}}	
	\prod_{nm}\exp{\biggl\{A_{nm}}
		\frac{\sin{\omega_0(n-m)t}}
		{\omega_0(n-m)}\biggr\}
	\biggr].
\end{eqnarray*}
\begin{figure}[t]
\begin{center}
\includegraphics[width=8.5cm]{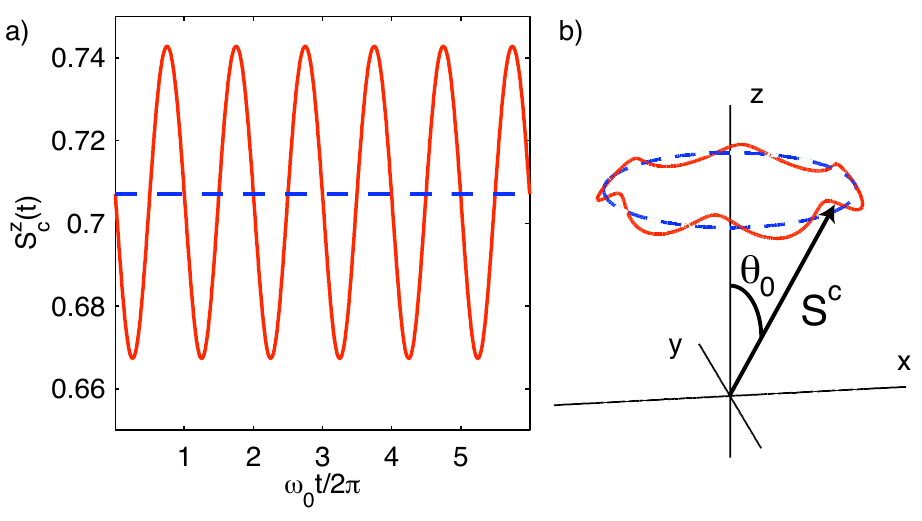}
\end{center}
\caption{(Color online) The different qualitative behavior of the spin $\bfS^c$ in presence of an applied ac voltage bias when the magnetizations of the leads are equal and FM-type (dashed), and AFM-type (solid) aligned.  a) The polar displacements of the spin dynamics. Here, we have defined spin-polarized density of states \cite{fransson2005} $N_{L\sigma/R\sigma}=N_0(1+\eta_\sigma p_{L/R})/2$ and $\eta_{\up/\down}=\pm1$, where $T_1N_0\sim 0.1$, $p_L=0.9$, $p_R=p_L$ (FM-type) and $p_R=-p_L$ (AFM-type), and $\omega_0=3$, which gives the induced field $g\mu_BB_\text{ind}\sim0.01eV_0$. b) The resulting generic spin motion on the unit sphere.}
\label{fig-precession}
\end{figure}
Generically, when a single spin is subjected to a uniform magnetic field, the spin azimuthally precesses with the Larmor frequency $\omega_L$. For the tunneling junction between ferromagnets, when a dc voltage bias is applied the spin precesses around an effective magnetic field $\bfB_\text{eff}=\bfB+\bfB^{(1)}+\bfB^{(2)}$ with a shifted Larmor frequency $\tilde\omega_L$. When an ac voltage bias is applied, Eq. (\ref{eq-angles}) shows that the spin also exhibits polar ($\theta$) modulations, whenever the magnetizations of the leads are non-equal and/or not in parallel configuration.  In particular,
when the external magnetic field $\bfB(t)=0$, the effective field in Eq. (\ref{eq-Seq}) becomes $\bfB_\text{eff}(t)=\bfB_\text{ind}^{(1)}(t)+\bfB_\text{ind}^{(2)}(t)$. Hence, we expect that the dynamics of the spin in absence of external magnetic fields is analogous to that in the presence of external magnetic fields. In Fig. \ref{fig-precession}a) we show the dynamics of the $z$-component of the spin, $S_z^c$, when the magnetizations of the leads are aligned in FM-type (dashed) and AFM-type (solid) configuration. In the plot, we only illustrate the first harmonic $\omega_0$, e.g. $n-m=1$ and $\sum_{n-m=1}A_{nm}$ constant, for a pure ac voltage, since the complete spin dynamics is expected to be very complicated.
Clearly, the polar angle $\theta$ exhibits a modulation with a dominant $\omega_0$ harmonic. 
The resulting dynamics of the spin can to much extent be compared with that of a rotating rigid top, as 
schematically illustrated  in Fig. \ref{fig-precession}b). Analogous to a classical spinning top, the spin wobbles along the polar direction in addition to the azimuthal rotations. 
The spin-polarized current leads to a full non-planar gyroscopic motion (nutations) of the spin much like that generated by applied torques on a mechanical top, i.e. magnetically induced nutations. 
Similar dynamics is expected to occur even for $S={1\over2}$ since in a spin coherent state the 
Schr\"odinger equation is essentially classical. The effect found in the present system is much similar to that of a single spin embedded in a Josephson junction~\cite{Zhu04}, although the origin is of different nature. In addition, the proposed system is experimentally more accessible because such challenges as extremely low temperatures and device engineering of a Josephson junction can be avoided here. 

In conclusion we find non-planar motion of a spin embedded in a tunnel junction between ferromagnets in presence of an alternating current, whenever the ferromagnets are unequally strong and/or in non-parallel alignment. The characteristic frequency of the polar angle variation equals that of the ac field, whereas the Larmor frequency of the spin precession scales as the squared ratio between the external and induced magnetic fields in a {\em dc} case. The induced magnetic field could be probed with superconducting quantum interference devices. 

We thank A. V. Balatsky and I. Martin for useful discussions. This work was carried out under the auspices of the NNSA of the US DOE  under Contract No. DE-AC52-06NA25396 and supported by the LDRD  and BES programs at Los Alamos.

\end{document}